# Imaging thermally fluctuating Néel vectors in van der Waals antiferromagnet NiPS$_3$

Youjin Lee[1,2], Chaebin Kim[1,2,‡], Suhan Son[1,2,¶], Jingyuan Cui[1,2†], Giung Park[1,2], Kai-Xuan Zhang[1,2], Siwon Oh[3], Hyeonsik Cheong[3], Armin Kleibert[4*], and Je-Geun Park[1,2*]

[1]Center for Quantum Materials, Seoul National University, Seoul, 08826, Republic of Korea
[2]Department of Physics and Astronomy & Institute of Applied Physics, Seoul National University, Seoul, 08826, Republic of Korea
[3]Department of Physics, Sogang University, Seoul, 04107, Republic of Korea
[4]Swiss Light Source, Paul Scherrer Institut, Villigen, 5232, Switzerland
‡ Present adress: School of Physics, Georgia Institute of Technology, Atlanta, Georgia, 30332, USA
¶ Present adress: Department of Physics, University of Michigan, Ann Arbor, Michigan, 48109, USA
† Present adress: Department of Physics & Astronomy, University of California, Riverside, California, 92521, USA

* Corresponding author: armin.kleibert@psi.ch & jgpark10@snu.ac.kr

**Abstract**

Studying antiferromagnetic domains is essential for fundamental physics and potential spintronics applications. Despite its importance, few systematic studies have been performed on van der Waals (vdW) antiferromagnets (AFMs) domains with high spatial resolutions, and direct probing of the Néel vectors remains challenging. In this work, we found a multidomain in vdW AFM NiPS$_3$, a material extensively investigated for its exotic magnetic exciton. We employed photoemission electron microscopy combined with the X-ray magnetic linear dichroism (XMLD-PEEM) to image the NiPS$_3$'s magnetic structure. The nanometer-spatial resolution of XMLD-PEEM allows us to determine local Néel vector orientations and discover thermally fluctuating Néel vectors that are independent of the crystal symmetry even at 65 K, well below T$_N$ of 155 K. We demonstrate a Ni ions' small in-plane orbital moment anisotropy is responsible for the weak magneto-crystalline anisotropy. The observed multidomain's thermal fluctuations may explain the broadening of magnetic exciton peaks at higher temperatures.

Two-dimensional magnetism has long been the subject of intensive and extensive studies, mainly theoretical investigations, until it was recently realized that some van der Waals magnetic materials could easily produce a single magnetic layer [1-5]. These new van der Waals magnets have since become an active field with immense potential [6]. The transition metal thiophosphates TMPS$_3$ are the first systems providing an experimental testbed for two-dimensional magnetism in the monolayer [5, 7]. These systems host antiferromagnetic order at low temperatures and exhibit different spin Hamiltonians depending on the transition metal ions [8].

NiPS$_3$ is a strongly correlated magnetic system and exhibits quantum-entangled magnetic excitons [9-11]. The observed excitons were assigned to the transition from two quantum entangled many-body states, the Zhang-Rice triplet to the Zhang-Rice singlet, through multiple spectroscopic probes [9], including optical absorption, photoluminescence (PL) and resonant inelastic X-ray scattering (RIXS). One of the most striking observations is the ultra-narrow linewidth, the full width of half maximum of 0.4 meV, and the sharp PL peak implies the coherent excitation of magnetic excitons in NiPS$_3$ [9]. We note that this linewidth of the exciton in NiPS$_3$ is narrower than those impurity peaks of a Ni defect, reportedly having about 7 meV [12]. It is also important to note that this exciton of NiPS$_3$ comes from a lattice of Ni ions, thus making it a great puzzle in the field. Simultaneously, polarization-dependent PL and optical linear dichroism experiments also found that the polarization of the magnetic excitons is locked to the local Néel vector [10]. These extraordinary observations indicate that the strong correlation between photons and spins makes NiPS$_3$ a promising magneto-optical device material.

The previous work also reported that the exciton forms a coherent state even far below T$_N$, the origin of which is still poorly understood. At the same time, NiPS$_3$ has a small in-plane magnetic anisotropy [13, 14]. Several key questions remain unanswered, such as its microscopic origin, its effect on the magnetic domain formation, and its role in the coherent state of the magnetic exciton. Thus, investigating domains and their formation is of great interest for fundamental research and designing magneto-optical devices based on the magnetic exciton of NiPS$_3$. Another equally exciting question concerns the nature of antiferromagnetic domain walls in NiPS$_3$. The dimensions and characteristics of these magnetic domain walls offer valuable insight into the properties of NiPS$_3$, such as magnetic stiffness and anisotropy [15, 16].

Despite advancements made over the past few years, the direct observation of antiferromagnetic domains in few-layered vdW crystals is still challenging primarily because of the experimental difficulties involved. Detecting the spin texture in those systems requires a high degree of spatial resolution and sensitivity to antiferromagnetic order. Historically, some of the first direct observations of antiferromagnetic domains were reported in bulk samples in the 1960s [17] and 70s [18-20] before antiferromagnetic domains were investigated in epitaxial thin films some 20 years later [21]. Several reports have been made on the antiferromagnetic domains of the transition metal thiophosphates TMPS$_3$ employing optical techniques. For example, the magnetic domains of MnPS$_3$ were observed using second harmonic generation (SHG) [22] as the broken inversion symmetry in the Néel type antiferromagnet of MnPS$_3$ gives rise to the two antiphase magnetic domains.

On the other hand, the optical linear dichroism (LD) technique was employed to observe an antiferromagnetic multidomain pattern in the few-layer FePS$_3$ [23, 24]. Although SHG and optical LD successfully probed antiferromagnetic domains in TMPS$_3$, new measurements with higher spatial resolutions are required to study spin texture in further detail as these optical probes have a spatial resolution of a few or tens of micrometer scales [22-25]. Another noteworthy point is that the symmetry breaking required for some techniques limits the optical measurements to a narrower case of the long-range magnetic order. For example, SHG cannot be used for NiPS$_3$ as the inversion symmetry is not broken in its antiferromagnetic order [26]. However, as done in FePS$_3$ [23, 24], optical LD measurements can be used for NiPS$_3$ as it has nonzero LD signals.

Here, we employ an X-ray spectromicroscopy technique to study the magnetic properties of NiPS$_3$ (Fig. 1A). The antiferromagnetic order induces temperature- and polarization-dependent modifications in the X-ray absorption (XA) spectra due to the optical selection rule [27, 28], resulting in the X-ray magnetic linear dichroism (XMLD). From the XMLD effect, we can detect the local Néel vectors directly and use X-ray photoemission electron microscopy (X-PEEM) to probe magnetic domains in NiPS$_3$ with sub-micrometer resolution and their dynamics to correlate them with their electronic structure. Another advantage of X-PEEM is the capability to probe the X-ray magnetic circular dichroism (XMCD) effect when using circularly polarized X-rays. In contrast to the XMLD effect, XMCD can detect only ferromagnetic order, i.e., local net magnetization. So, XMCD can be used to identify domain walls where uncompensated local magnetic moments lead to net nonzero magnetization [29].

By combining local X-ray absorption spectroscopy with XMLD- and XMCD-PEEM imaging with atomic force microscopy (AFM), optical microscopy, multiplet calculation, and Monte-Carlo simulation, we demonstrate that the weak magneto-crystalline anisotropy in NiPS$_3$ is due to a small in-plane magnetic orbital moment anisotropy of the Ni ion. Furthermore, the weak anisotropy induces thermally fluctuating antiferromagnetic domains with their local Néel vectors pointing in arbitrary directions. These results show that magnetic anisotropy related to the multidomain structure has to be considered for the design of magneto-optical devices where excitons are strongly correlated with spins like in NiPS$_3$.

The bulk NiPS$_3$ has the monoclinic point group $C_{2h}$. The Ni atoms are arranged in a hexagonal lattice, surrounded by six S atoms with a trigonal symmetry. It is to be noted that although less recognized, the Ni honeycomb lattice has two distinct Ni-Ni bond distances of 3.3551 and 3.3543 Å (see Fig. S1). This breaks the three-fold rotation of an otherwise ideal honeycomb lattice. Below the bulk Néel temperature of $T_N$ = 155 K, the spins are aligned predominantly along the *a*-axis with a small out-of-plane component of 15.5° towards the *c*-axis [30]. The spin order forms ferromagnetic zig-zag chains that are antiferromagnetically coupled along the *b*-axis (Fig. 1B). Each ferromagnetic chain has net magnetization $M_1$ and $M_2$, and the Néel vector *L* can be defined as $L = M_1 - M_2$ with the Néel vector parallel to the *a*-axis in the bulk (Fig. 1C).

Fig. 1E shows XMLD-PEEM images obtained at the Ni L$_3$-edge at 70 K, readily revealing antiferromagnetic domains. The width of the dark stripe-like domain in Fig. 1E is estimated to be about 200 nm, indicated by the two arrows. To the best of our knowledge, it is the first report of vdW AF material NiPS$_3$, which has been only possible because of the high spatial resolution of our X-PEEM technique. Here, we focus on stripe-like domains showing clear magnetic contrast and thermal fluctuations, and the domain patterns visibly vary across flakes (Fig. S2). We confirmed the flatness of the examined region in the samples using a separate atomic force microscope measurement, which excludes the surface morphology as the source of the observed domains (Fig. S3). Fig. 1D shows the X-ray propagation vector *k* and both out-of-plane and in-plane electric field directions together with the crystalline axes of the NiPS$_3$ sample. The crystalline axes were determined by combining our polarized Raman spectroscopy experiments (Fig. S4) and the morphology of the flake (See supplementary for more detail). We note that X-PEEM images taken with circular polarized X-rays (C+ and C-) did not reveal any detectable signal of the domain walls in the same flake. This absence is attributed to the narrow width of the domain walls below the resolution of X-PEEM, a point that will be further discussed later. In Fig. S5, we show the XMCD contrast image acquired at photon energy at the Ni L$_3$-edge at 68 K.

To confirm the magnetic origin of the domain pattern observed in Fig. 1E, we increased the temperature to 285 K, well above the bulk Néel temperature of 155 K (Fig. 2). From this, we observed that the XMLD contrast in Fig. 2 visible in the 70 K data decreases with increasing temperature. When the temperature returns to 70 K, the magnetic domains reappear but now have a different domain pattern than before the thermal cycle (Fig. 2D). It is a clear sign of thermal cycling effects on the magnetic domains.

      Figures 3A and 3B show the experimental XA spectra of domain A as in Fig. 3C, which were taken below and above $T_N$ (70 and 285 K). As one can see, the linear dichroic signal at 285 K decreases compared to the antiferromagnetic state at 70 K. In the antiferromagnetic state, the long-range magnetic order is expected to produce a substantial temperature-dependent linear dichroic signal with the XMLD effect [21, 31]. Therefore, the enhanced linear dichroic signal below the Néel temperature originates from the long-range magnetic order. This magnetic linear dichroism from the long-range order vanishes in the paramagnetic state, leaving only weak linear dichroism that arises from orbital anisotropy and short-range magnetic order.

We verify the origin of the linear dichroism through multiplet ligand field calculations [32] (Fig. S6). Note that linear dichroism is present at all temperatures due to trigonal distortion. So first, we fit the XA and XLD spectra taken at 285 K using the parameter of the trigonal symmetry $D_{3d}$ [8, 33]. After that, the long-range magnetic order is included in our calculations through the exchange field to analyze the data below $T_N$. As shown at the bottom of Figs. 3A and B, the calculated spectra reproduce the experimental data well. A weak linear dichroic signal above $T_N$ can also originate from the short-range magnetic order [34]. In the case of NiPS$_3$, the respective small remaining XMLD contrast is present in an asymmetric XMLD asymmetry image taken at 285 K (Fig. 2C). Usually, the nearest spin-spin correlation can be ignored in the XMLD analysis due to the small contribution compared to the long-range antiferromagnetic order. However, enhanced short-range order is the critical feature of two-dimensional magnetism like NiPS$_3$, resulting in a difference between the susceptibility maximum temperature and $T_N$ [35]. For example, the magnetic susceptibility of bulk NiPS$_3$ shows a broad maximum at as high as 300 K in the paramagnetic state, indicating the presence of a strong short-range magnetic order well above the Néel temperature (Fig. S7). In our multiplet calculations, the small XMLD effect of the short-range magnetic order is captured by the orbital parameters. This is because the short-range spin correlation cannot be described by a mean-field type of theory, which is the exchange field term in our calculation [34].

We can further determine the Néel vector orientation in each domain by analyzing the experimental XMLD spectra using the same multiplet calculation method while varying the direction of the exchange field (Fig. 3C-E). The Néel vector orientations in different domains are depicted in Fig. 3C by double-sided arrows. For example, the Néel vector in the A domain is parallel to the *a*-axis, whereas the Néel vectors were rotated by 10º ($\pm$ 9.3º) from the *a*-axis in the B, D and E domains and 20º ($\pm$ 7.4º) from the *a*-axis in the C domain, respectively. The error is estimated from the uncertainty of the peak's height in the Voigtian functions that are used to fit the raw data. We note that the determined Néel vector orientations do not show any sign the three-fold rotation symmetry. As mentioned before, this observation is consistent with the fact that the Ni honeycomb lattice lacks the three-fold rotation symmetry. Closely related, multiple Néel vectors were also observed from the magnetic field experiment on the excitons in bulk NiPS$_3$ [10]. When the external magnetic field was applied along the direction that is 120º rotated from the *a*-axis, the Néel vector orientation was also found to rotate accordingly. Interestingly, with an increasing external magnetic field, the Néel vector is observed to rotate continuously, not at 60º intervals, which seems consistent with our data. These Néel vector orientations originate from the small magnetic anisotropy, as discussed below.

Of further interest is that our experiment reveals thermal fluctuations of the magnetic domains during cooling (Fig. 4 & Movie S1 in Supplementary Information). For example, the B and C domains merged into the A domain with decreasing temperature, where the rotated Néel vectors turn parallel to the *a*-axis. The thermal fluctuation of domains seen in our data indicates that the magnetic moments of NiPS$_3$ are still mobile, not firmly pinned, even down to 65 K, corresponding to $\tau$=0.41 of the reduced temperature ($\tau$=T/T$_N$). Traditionally, the formation of antiferromagnetic multidomain states originates from several reasons, such as crystalline twin structures [36, 37], and magnetoelastic effects, including the destressing field from the substrate [38, 39]. Here, the thermally fluctuating domains exclude a structural origin of the observed domains. The domains from the crystalline twin structure and clamping effects due to the substrates would be expected to show temperature-independent domain morphologies because magnetostriction-induced strain hinders the growth of the domains and fluctuations [21, 40]. Another interesting point is its relevance to the magnetic coherence of excitons in NiPS3. In the previous PL experiment on NiPS3, the spin-correlated exciton PL peak width gets decreased with reducing temperature and approached the ultra-narrow width below 50 K [9]. An important point is that an electric dipole moment emerges spontaneously below the antiferromagnetic transition [9]. This electric dipole points along the *b*-axis, which determines the extreme polarization of the PL peak, another sign of the strong coupling among the exciton, the spin moments, and the electric dipole. Under such situations, mobile magnetic domains can easily affect the lifetime of the exciton and so its linewidth. Therefore, the broadening of the exciton peak can be explained by the thermal instability of the Néel vectors above 50 K, which is observed in XMLD-PEEM images.

Next, we want to discuss the weak magneto-crystalline anisotropy NiPS3, which would allow the thermal energy of our experimental temperature regime to rotate the Néel vector within the *ab*-plane. Two recent neutron scattering experiments estimated the value of 0.01 meV per Ni ion for an easy axis anisotropy along the *a*-axis [14, 41]. We simulated the magnetic domains to identify the weak magnetic anisotropy as the central origin of the observed multidomain state. The spin configuration was simulated using Metropolis-adjusted Langevin algorithms [42] and the spin Hamiltonian as determined from inelastic neutron scattering experiments [14]. We then converted the spin configuration to an XMLD contrast map using the equation of XMLD intensity, $I(\theta) = a + b(3cos^2\theta - 1)\langle M^2 \rangle$ $(eq. 1)$, where $\theta$ is the angle between the incident *E*-polarization of the X-rays and the Néel vector [31]. These theoretical studies successfully reproduce the orientation of different local Néel vectors from the spin Hamiltonian of weak magnetic anisotropy (Fig. S8). As in the experiment, the multidomain state in the calculation merges into a single domain state where all Néel vectors are aligned along the *a*-axis as the temperature approaches zero.

While we can understand the observed magnetic domain dynamics based on the weak magnetic anisotropy, the XA spectra provide further insight into the latter's origin. Our multiplet simulations fitted to the experimental XA spectra yield a slightly anisotropic in-plane orbital moment of $L_a = 0.2630 \pm 0.0002\,\mu_B/Ni$ and $L_b = 0.2557 \pm 0.0003\,\mu_B/Ni$. We think that their small anisotropy within an *ab* plane leads to weak magnetocrystalline anisotropy [43, 44]. We further estimate the in-plane magnetocrystalline anisotropy to be 0.167 meV/*Ni*, using the spin-orbit coupling constant and the orbital moment anisotropy obtained from the multiplet simulations. Our estimate of the magnetic anisotropy can be converted to $K = 4.35 \times 10^6$ erg/cm³.

The XMCD images calculated from (I+ - I-)/(I+ + I-) do not show possibly uncompensated magnetic moments, i.e. local ferromagnetic order, in the domain wall (Fig. S5). This indicates that the domain wall size should be smaller than the spatial resolution of X-PEEM, which we estimate here to be 45 nm. The domain wall width is usually determined by competition between the exchange stiffness ($A_{ex}$) and the anisotropy ($K$) [15]. When Néel vectors get rotated by 30° with respect to each other, as in the B and C domains, we can obtain the expression for the following domain wall width, $w = \frac{\pi}{6}\sqrt{A_{ex}/K}$ $(eq. 2)$. When we use the exchange parameter from the spin Hamiltonian [14] with $A_{ex} = 27.2$ meV/nm and $K = 2.71$ meV/nm³ determined above, the calculated domain wall width turns out to be 1.66 nm, which explains why we failed to observe in our experiments.

To summarize, we observe magnetic domains with different Néel vector orientations using XMLD-PEEM. The high spatial resolution of the X-ray imaging technique enables us to determine domain widths down to 200 nm in NiPS$_3$ in real space. At the same time, the analysis of the XMLD data allows us direct access to local Néel vectors. The temperature-dependent XMLD intensities and thermally fluctuating domains manifest their magnetic origins, excluding domains from crystalline twin structures. Our analysis of the temperature-dependent domain morphology demonstrates a weak magnetic anisotropy of NiPS$_3$. We also found that the weak magnetic anisotropy originates from the small anisotropy in the magnetic orbital moments within the *ab*-plane. Therefore, Néel vectors can take arbitrary in-plane orientations and are unstable against thermal fluctuations. These findings have an exciting implication for the finite coherence of the magnetic excitons reported in this system. Our work emphasizes the importance of XMLD-PEEM as an ideal technique for probing vdW antiferromagnetic domains, revealing the fundamental physics of magnetic anisotropy in their formation and dynamics.

# References

(1) Park, J.-G. Opportunities and challenges of 2D magnetic van der Waals materials: magnetic graphene? *J. Phys. Condens. Matter* **2016**, *28* (30), 301001. DOI: 10.1088/0953-8984/28/30/301001.

(2) Kuo, C.-T.; Neumann, M.; Balamurugan, K.; Park, H. J.; Kang, S.; Shiu, H. W.; Kang, J. H.; Hong, B. H.; Han, M.; Noh, T. W. Exfoliation and Raman spectroscopic fingerprint of few-layer $NiPS_3$ van der Waals crystals. *Sci. Rep.* **2016**, *6* (1), 1-10. DOI: 10.1038/srep20904 (2016).

(3) Gong, C.; Li, L.; Li, Z.; Ji, H.; Stern, A.; Xia, Y.; Cao, T.; Bao, W.; Wang, C.; Wang, Y.; et al. Discovery of intrinsic ferromagnetism in two-dimensional van der Waals crystals. *Nature* **2017**, *546* (7657), 265-269. DOI: 10.1038/nature22060.

(4) Huang, B.; Clark, G.; Navarro-Moratalla, E.; Klein, D. R.; Cheng, R.; Seyler, K. L.; Zhong, D.; Schmidgall, E.; McGuire, M. A.; Cobden, D. H.; et al. Layer-dependent ferromagnetism in a van der Waals crystal down to the monolayer limit. *Nature* **2017**, *546* (7657), 270-273. DOI: 10.1038/nature22391.

(5) Lee, J. U.; Lee, S.; Ryoo, J. H.; Kang, S.; Kim, T. Y.; Kim, P.; Park, C. H.; Park, J. G.; Cheong, H. Ising-Type Magnetic Ordering in Atomically Thin $FePS_3$. *Nano Lett.* **2016**, *16* (12), 7433-7438. DOI: 10.1021/acs.nanolett.6b03052.

(6) Burch, K. S.; Mandrus, D.; Park, J.-G. Magnetism in two-dimensional van der Waals materials. *Nature* **2018**, *563* (7729), 47-52. DOI: 10.1038/s41586-018-0631-z.

(7) Lee, S.; Choi, K.-Y.; Lee, S.; Park, B. H.; Park, J.-G. Tunneling transport of mono- and few-layers magnetic van der Waals $MnPS_3$. *APL Mater.* **2016**, *4* (8). DOI: 10.1063/1.4961211 (acccessed 1/10/2024).

(8) Joy, P. A.; Vasudevan, S. Magnetism in the layered transition-metal thiophosphates $MPS_3$ (M=Mn, Fe, and Ni). *Phys. Rev. B.* **1992**, *46* (9), 5425-5433. DOI: 10.1103/physrevb.46.5425.

(9) Kang, S.; Kim, K.; Kim, B. H.; Kim, J.; Sim, K. I.; Lee, J.-U.; Lee, S.; Park, K.; Yun, S.; Kim, T. Coherent many-body exciton in van der Waals antiferromagnet $NiPS_3$. *Nature* **2020**, *583* (7818), 785-789. DOI: 10.1038/s41586-020-2520-5.

(10) Wang, X.; Cao, J.; Lu, Z.; Cohen, A.; Kitadai, H.; Li, T.; Tan, Q.; Wilson, M.; Lui, C. H.; Smirnov, D. Spin-induced linear polarization of photoluminescence in antiferromagnetic van der Waals crystals. *Nat. Mater.* **2021**, *20* (7), 964-970. DOI: 10.1038/s41563-021-00968-7.

(11) Hwangbo, K.; Zhang, Q.; Jiang, Q.; Wang, Y.; Fonseca, J.; Wang, C.; Diederich, G. M.; Gamelin, D. R.; Xiao, D.; Chu, J.-H. Highly anisotropic excitons and multiple phonon bound states in a van der Waals antiferromagnetic insulator. *Nat. Nanotechnol.* **2021**, *16* (6), 655-660. DOI: 10.1038/s41565-021-00873-9.

(12) Ho, C.-H.; Hsu, T.-Y.; Muhimmah, L. C. The band-edge excitons observed in few-layer $NiPS_3$. *npj 2D Mater. Appl.* **2021**, *5* (1), 8. DOI: 10.1038/s41699-020-00188-8.

(13) Kim, T. Y.; Park, C.-H. Magnetic anisotropy and magnetic ordering of transition-metal phosphorus trisulfides. *Nano Lett.* **2021**, (21), 10114-10121. DOI: 10.1021/acs.nanolett.1c03992.

(14) Scheie, A.; Park, P.; Villanova, J. W.; Granroth, G. E.; Sarkis, C. L.; Zhang, H.; Stone, M. B.; Park, J.-G.; Okamoto, S.; Berlijn, T.; et al. Spin wave Hamiltonian and anomalous scattering in $NiPS_3$. *Phys. Rev. B* **2023**, *108* (10),

104402. DOI: 10.1103/PhysRevB.108.104402.

(15) Jani, H.; Lin, J.-C.; Chen, J.; Harrison, J.; Maccherozzi, F.; Schad, J.; Prakash, S.; Eom, C.-B.; Ariando, A.; Venkatesan, T.; et al. Antiferromagnetic half-skyrmions and bimerons at room temperature. *Nature* **2021**, *590* (7844), 74-79. DOI: 10.1038/s41586-021-03219-6.

(16) Parkin, S.; Yang, S.-H. Memory on the racetrack. *Nat. Nanotechnol.* **2015**, *10* (3), 195-198. DOI: 10.1038/nnano.2015.41.

(17) Roth, W. Neutron and optical studies of domains in NiO. *J. Appl. Phys.* **1960**, *31* (11), 2000-2011. DOI: 10.1063/1.1735486.

(18) Safa, M.; Tanner, B. K. Behaviour of antiferromagnetic domains in $KNiF_3$. *Physica B+C* **1977**, *86*, 1347-1348. DOI: 10.1016/0378-4363(77)90906-8.

(19) Nakahigashi, K.; Fukuoka, N.; Shimomura, Y. Crystal structure of antiferromagnetic NiO determined by X-ray topography. *J. Phys. Soc. Jpn.* **1975**, *38* (6), 1634-1640. DOI: 10.1143/JPSJ.38.1634.

(20) Safa, M.; Tanner, B. K. Antiferromagnetic domain wall motion in $KNiF_3$ and $KCoF_3$ observed by X-ray synchrotron topography. *Philos. mag., B* **1978**, *37* (6), 739-750. DOI: 10.1080/01418637808225652.

(21) Scholl, A.; Stöhr, J.; Lüning, J.; Seo, J. W.; Fompeyrine, J.; Siegwart, H.; Locquet, J.-P.; Nolting, F.; Anders, S.; Fullerton, E. Observation of antiferromagnetic domains in epitaxial thin films. *Science* **2000**, *287* (5455), 1014-1016. DOI: 10.1126/science.287.5455.1014.

(22) Ni, Z.; Zhang, H.; Hopper, D. A.; Haglund, A. V.; Huang, N.; Jariwala, D.; Bassett, L. C.; Mandrus, D. G.; Mele, E. J.; Kane, C. L. Direct Imaging of Antiferromagnetic Domains and Anomalous Layer-Dependent Mirror Symmetry Breaking in Atomically Thin $MnPS_3$. *Phys. Rev. Lett.* **2021**, *127* (18), 187201. DOI: 10.1103/PhysRevLett.127.187201.

(23) Ni, Z.; Huang, N.; Haglund, A. V.; Mandrus, D. G.; Wu, L. Observation of Giant Surface Second-Harmonic Generation Coupled to Nematic Orders in the van der Waals Antiferromagnet $FePS_3$. *Nano Lett.* **2022**, *22* (8), 3283-3288. DOI: 10.1021/acs.nanolett.2c00212.

(24) Zhang, Q.; Hwangbo, K.; Wang, C.; Jiang, Q.; Chu, J.-H.; Wen, H.; Xiao, D.; Xu, X. Observation of giant optical linear dichroism in a zigzag antiferromagnet $FePS_3$. *Nano Lett.* **2021**, *21* (16), 6938-6945. DOI: 10.1021/acs.nanolett.1c02188.

(25) Kim, D. S.; Huang, D.; Guo, C.; Li, K.; Rocca, D.; Gao, F. Y.; Choe, J.; Lujan, D.; Wu, T. H.; Lin, K. H. Anisotropic Excitons Reveal Local Spin Chain Directions in a van der Waals Antiferromagnet. *Adv. Mater.* **2023**, 2206585. DOI: 10.1002/adma.202206585.

(26) Chu, H.; Roh, C. J.; Island, J. O.; Li, C.; Lee, S.; Chen, J.; Park, J.-G.; Young, A. F.; Lee, J. S.; Hsieh, D. Linear Magnetoelectric Phase in Ultrathin $MnPS_3$ Probed by Optical Second Harmonic Generation. *Phys. Rev. Lett.* **2020**, *124* (2), 027601. DOI: 10.1103/PhysRevLett.124.027601.

(27) Thole, B. T.; Van der Laan, G.; Sawatzky, G. A. Strong magnetic dichroism predicted in the $M_{4,5}$ X-ray absorption spectra of magnetic rare-earth materials. *Phys. Rev. Lett.* **1985**, *55* (19), 2086. DOI:

10.1103/PhysRevLett.55.2086.

(28) van der Laan, G.; Thole, B. T. Strong magnetic x-ray dichroism in 2p absorption spectra of 3d transition-metal ions. *Phys. Rev. B.* **1991**, *43* (16), 13401-13411. DOI: 10.1103/physrevb.43.13401.

(29) Sass, P. M.; Ge, W.; Yan, J.; Obeysekera, D.; Yang, J.; Wu, W. Magnetic imaging of domain walls in the antiferromagnetic topological insulator $MnBi_2Te_4$. *Nano Lett.* **2020**, *20* (4), 2609-2614. DOI: 10.1021/acs.nanolett.0c00114.

(30) Wildes, A. R.; Simonet, V.; Ressouche, E.; McIntyre, G. J.; Avdeev, M.; Suard, E.; Kimber, S. A. J.; Lançon, D.; Pepe, G.; Moubaraki, B.; et al. Magnetic structure of the quasi-two-dimensional antiferromagnet $NiPS_3$. *Phys. Rev. B.* **2015**, *92* (22), 224408. DOI: 10.1103/PhysRevB.92.224408.

(31) Stöhr, J.; Scholl, A.; Regan, T.; Anders, S.; Lüning, J.; Scheinfein, M.; Padmore, H.; White, R. Images of the antiferromagnetic structure of a NiO (100) surface by means of X-ray magnetic linear dichroism spectromicroscopy. *Phys. Rev. Lett.* **1999**, *83* (9), 1862. DOI: 10.1103/PhysRevLett.83.1862.

(32) Haverkort, M. W.; Zwierzycki, M.; Andersen, O. K. Multiplet ligand-field theory using Wannier orbitals. *Phys. Rev. B.* **2012**, *85* (16), 165113. DOI: 10.1103/PhysRevB.85.165113.

(33) Lee, Y.; Son, S.; Kim, C.; Kang, S.; Shen, J.; Kenzelmann, M.; Delley, B.; Savchenko, T.; Parchenko, S.; Na, W. Giant magnetic anisotropy in the atomically thin van der Waals antiferromagnet $FePS_3$. *Adv. Electron. Mater.* **2022**, 2200650. DOI: 10.1002/aelm.202200650.

(34) Alders, D.; Vogei, J.; Levelut, C.; Peacor, S.; Hibma, T.; Sacchi, M.; Tjeng, L.; Chen, C.; Van der Laan, G.; Thole, B. Magnetic x-ray dichroism study of the nearest-neighbor spin-spin correlation function and long-range magnetic order parameter in antiferromagnetic NiO. *EPL* **1995**, *32* (3), 259. DOI: 10.1209/0295-5075/32/3/012.

(35) Lines, M. E. Magnetism in two dimensions. *J. Appl. Phys.* **1969**, *40* (3), 1352-1358. DOI: /10.1063/1.1657665.

(36) Reimers, S.; Kriegner, D.; Gomonay, O.; Carbone, D.; Krizek, F.; Novák, V.; Campion, R. P.; Maccherozzi, F.; Björling, A.; Amin, O. J.; et al. Defect-driven antiferromagnetic domain walls in CuMnAs films. *Nat. Commun.* **2022**, *13* (1), 724. DOI: 10.1038/s41467-022-28311-x.

(37) Krizek, F.; Reimers, S.; Kašpar, Z.; Marmodoro, A.; Michalička, J.; Man, O.; Edström, A.; Amin, O. J.; Edmonds, K. W.; Campion, R. P.; et al. Atomically sharp domain walls in an antiferromagnet. *Sci. Adv.* **2022**, *8* (13), eabn3535. DOI: 10.1126/sciadv.abn3535.

(38) Baldrati, L.; Ross, A.; Niizeki, T.; Schneider, C.; Ramos, R.; Cramer, J.; Gomonay, O.; Filianina, M.; Savchenko, T.; Heinze, D.; et al. Full angular dependence of the spin Hall and ordinary magnetoresistance in epitaxial antiferromagnetic NiO(001)/Pt thin films. *Phys. Rev. B.* **2018**, *98* (2), 024422. DOI: 10.1103/PhysRevB.98.024422.

(39) Wittmann, A.; Gomonay, O.; Litzius, K.; Kaczmarek, A.; Kossak, A. E.; Wolf, D.; Lubk, A.; Johnson, T. N.; Tremsina, E. A.; Churikova, A. Role of substrate clamping on anisotropy and domain structure in the canted antiferromagnet α− $Fe_2O_3$. *Phys. Rev. B.* **2022**, *106* (22), 224419. DOI: 10.1103/PhysRevB.106.224419.

(40) Xu, J.; Zhou, C.; Jia, M.; Shi, D.; Liu, C.; Chen, H.; Chen, G.; Zhang, G.; Liang, Y.; Li, J. Imaging antiferromagnetic domains in nickel oxide thin films by optical birefringence effect. *Phys. Rev. B.* **2019**, *100* (13), 134413. DOI:

10.1103/PhysRevB.100.134413.

(41) Wildes, A.; Stewart, J.; Le, M.; Ewings, R.; Rule, K.; Deng, G.; Anand, K. Magnetic dynamics of $NiPS_3$. *Phys. Rev. B.* **2022**, *106* (17), 174422.

(42) Skubic, B.; Hellsvik, J.; Nordström, L.; Eriksson, O. A method for atomistic spin dynamics simulations: implementation and examples. *J. Phys.: Condens. Matter* **2008**, *20* (31), 315203. DOI: 10.1088/0953-8984/20/31/315203.

(43) Bruno, P. Tight-binding approach to the orbital magnetic moment and magnetocrystalline anisotropy of transition-metal monolayers. *Phys. Rev. B.* **1989**, *39* (1), 865. DOI: 10.1103/PhysRevB.39.865.

(44) van der Laan, G. Microscopic origin of magnetocrystalline anisotropy in transition metal thin films. *J. Phys. Condens. Matter* **1998**, *10* (14), 3239. DOI: 10.1088/0953-8984/10/14/012.

**Supporting information:** Methods, additional experimental data (antiferromagnetic domains in another NiPS$_3$ flake, AFM profile, polarization-dependent Raman, magnetic susceptibility and XMCD asymmetric image), and theoretical data (ligand multiplet calculations and Monte-Carlo simulations). Movie S1 of fluctuating antiferromagnetic domains

**Acknowledgement**

We are grateful to Gun Sang Jeon for the helpful discussion and Michel Kenzelmann for the kind help at the early stage of this project.

**Funding** This work was supported by the Leading Researcher Program of the National Research Foundation of Korea (Grant No. 2020R1A3B2079375). This work was also partly supported by the Ministry of Education through the core center program (2021R1A6C101B418).

**Author contributions** J.-G.P. and A.K. supervised this project. Y.L. prepared the sample. Y.L., S.S., J.C., G.P., and A.K. performed the X-PEEM experiment. Y.L. analyzed the experimental data. C.K. performed the Monte-Carlo simulation. S.O. and H.C. conducted the Raman experiment. Y.L., C.K., K.Z., A.K., and J.-G.P. wrote the manuscript with inputs from all authors.

**Competing interests:** The authors declare no competing interests.

**Data and materials availability:** All data are presented in the main text or the supplementary materials.

# Figures

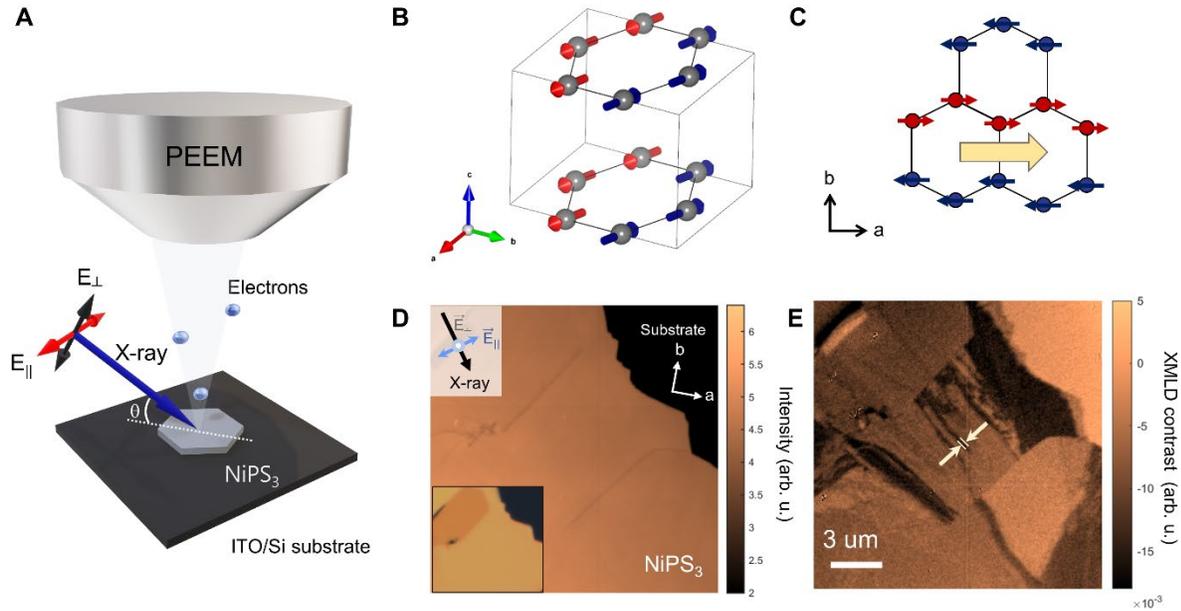

**Fig. 1. Scheme of X-PEEM experiment and NiPS₃ magnetic structure (**A) Schematic of the X-PEEM experiment with two types of linearly polarized X-rays, $E_\parallel$ and $E_\perp$, impinged on the NiPS₃ flake at a grazing angle θ = 16º. (B) Magnetic structure of NiPS₃. (C) In-plane view of the magnetic structure of NiPS₃ with the Néel vector (yellow arrow). (D) The normalized X-PEEM image. The inset is the optical microscope (OM) image of the same sample. The flake is flat and 48 nm thick, except for the orange-colored region in the OM image. Information on the sample and X-ray geometries are demonstrated as arrows. The black arrow indicates the direction of the incoming X-ray, and the blue double arrow indicates its in-plane electric field polarization. The out-of-plane electric field polarization is shown by the sky-blue dot symbol. (E) The corresponding asymmetric XMLD image. The normalized and asymmetric XMLD images are measured at a photon energy of 850.2 eV and 70 K.

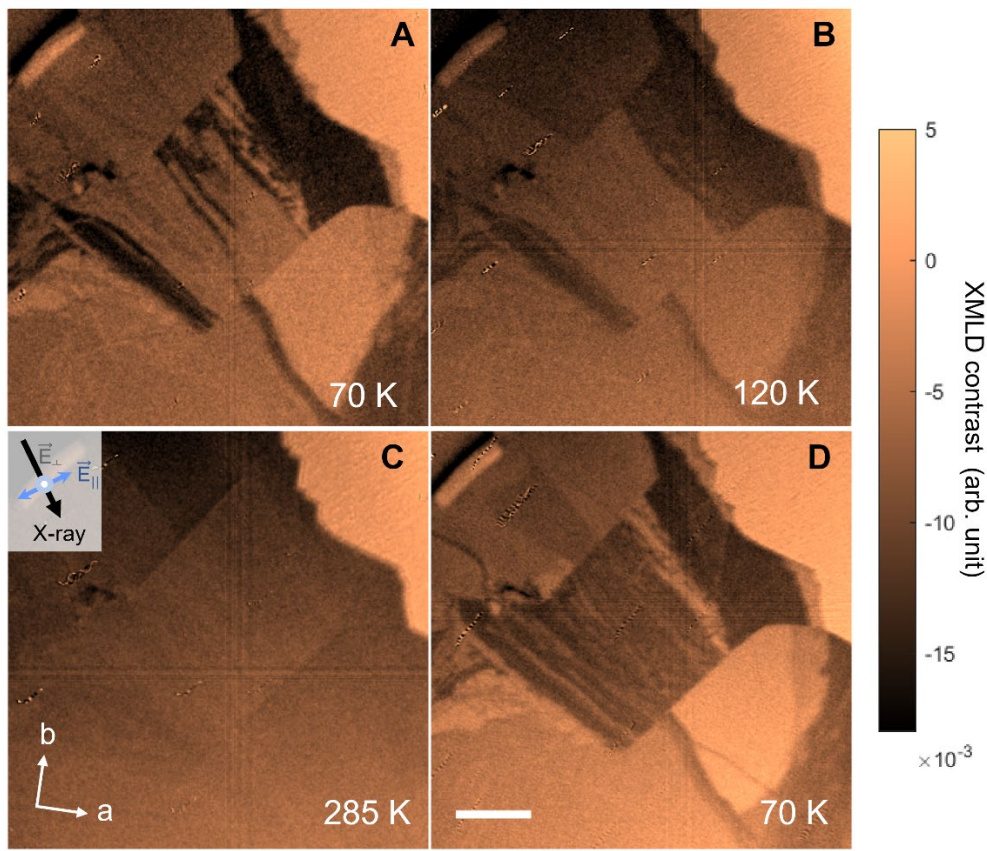

**Fig. 2. Temperature dependence of antiferromagnetic domains** (A-D) XMLD asymmetric images as a function of temperature. The scale bar in (D) is for 3 μm.

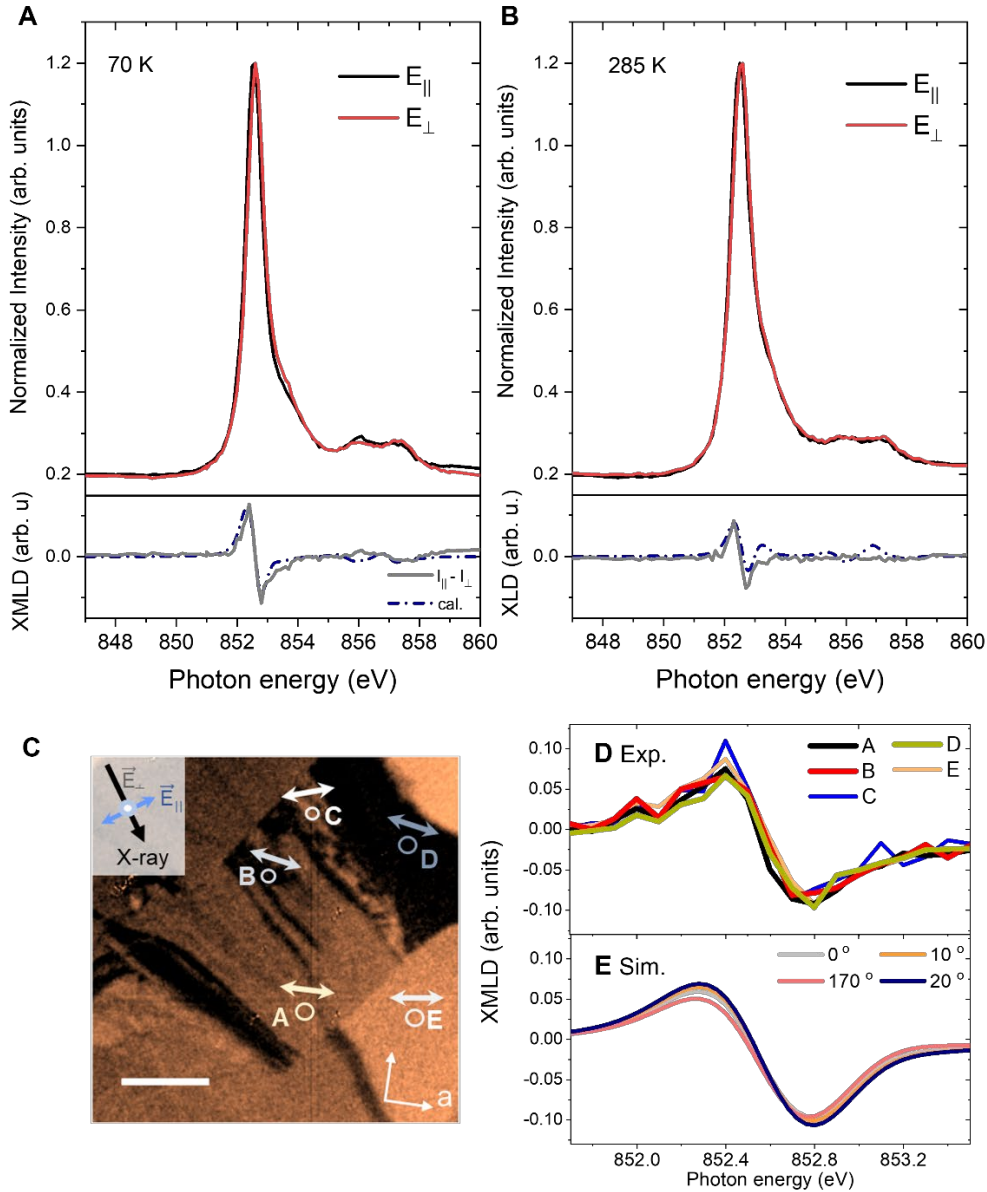

**Fig. 3. X-ray absorption spectra and multiplet calculations** The experimental XA spectra were taken at (A) 70 K (T < $T_N$) and (B) 285 K (T > $T_N$). Experimental (solid grey line) and calculated (blue dotted line) XMLD spectra are plotted at the bottom. (C) Antiferromagnetic domains with different XMLD contrasts are notated as A, B, C, D, and E. Double arrows in each domain indicate the Néel vector axis inferred from the result of (E). The scale bar is for 3 μm. (D) Experimental XMLD spectra from the magnetic domains A-E. (E) Calculated XMLD spectra with different Néel vector axes. The assigned Néel vector axes from the calculation are parallel to the *a*-axis in the A domain, 10º (± 9.3º) rotated from the *a*-axis in the B, D and E domains, and 20º (± 7.4º) rotated from the *a*-axis in the C domain.

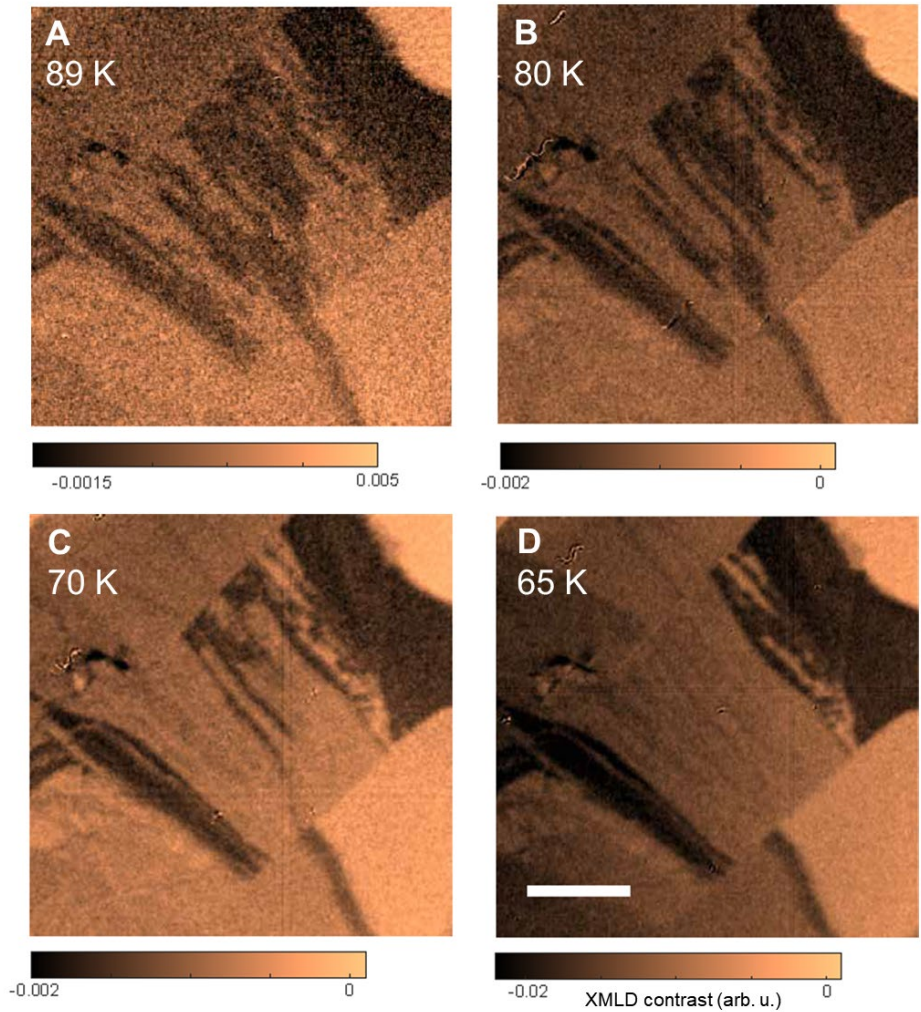

**Fig. 4. Thermal fluctuations of antiferromagnetic domains** (A-D) Thermally fluctuating antiferromagnetic domains in XMLD asymmetric images. To emphasize the domain fluctuation, the contrast ranges were adjusted in images of each temperature. The scale bar is for 3 μm.